\documentclass[runningheads]{llncs}
\usepackage[T1]{fontenc}
\usepackage{amsmath}
\usepackage{graphicx}
\usepackage{enumitem}
\usepackage{booktabs}
\usepackage{hyperref}
\begin{document}
\title{Decoding Student Dialogue: A Multi-Dimensional Comparison and Bias Analysis of Large Language Models as Annotation Tools}
\titlerunning{Decoding Student Dialogue: Large Language Models as Annotation Tools}
%
 \author{Jie Cao\inst{1}\and
Zhanxin Hao\inst{2} \and
Jifan Yu\inst{2}}
 \authorrunning{Cao et al.}
\institute{The University of North Carolina at Chapel Hill, North Carolina, USA\and
 Tsinghua University, China\\
 \email{jiecao@unc.edu}\\
 \email{\{zhanxin\_hao,yujifan\}@tsinghua.edu.cn}}
\maketitle              
\begin{abstract}
Educational dialogue is critical for decoding student learning processes, yet manual annotation remains time-consuming. This study evaluates the efficacy of GPT-5.2 and Gemini-3 using three prompting strategies (few-shot, single-agent, and multi-agent reflection) across diverse subjects, educational levels, and four coding dimensions. Results indicate that while multi-agent prompting achieved the highest accuracy, the results did not reach statistical significance. Accuracy proved highly context-dependent, with significantly higher performance in K-12 datasets compared to university-level data, alongside disciplinary variations within the same educational level. Performance peaked in the affective dimension but remained lowest in the cognitive dimension. Furthermore, analysis revealed four bias patterns: (1) Gemini-3 exhibited a consistent optimistic bias in the affective dimension across all subjects; (2) the cognitive dimension displayed domain-specific directional bias, characterized by systematic underestimation in Mathematics versus overestimation in Psychology; (3) both models are more prone to overestimation than underestimation within the meta-cognitive dimension; and (4) behavioral categories such as question, negotiation, and statements were frequently misclassified. These results underscore the need for context-sensitive deployment and targeted mitigation of directional biases in automated annotation.

\end{abstract}

\section{Introduction}
Educational dialogue is a fundamental vehicle for knowledge exchange, and its quality significantly influences student learning outcomes \cite{muhonen2018quality,howe2019teacher,hao_mapping_2026}. While analyzing these interactions through established frameworks provides deep insights into the learning process \cite{hennessy2016developing}, manual coding remains prohibitively time-consuming \cite{long2024evaluating}. This analytical bottleneck is exacerbated by the proliferation of Large Language Models (LLMs) in education, which has made student-AI interactions ubiquitous \cite{kasneci2023chatgpt}. Consequently, educators now face a massive volume of dialogue data that needs to be analyzed to understand student engagement and inform instruction, necessitating robust, automated approaches.

Recent studies have already explored the ``LLM-as-a-judge'' paradigm for automated educational annotation, demonstrating significant potential \cite{long2024evaluating,he2025automated,nguyen2025applying,jiang2025uncovering}. However, reported accuracies vary widely (ranging from 0.6 to 0.9), likely due to differences in prompt engineering, model capabilities, contextual factors, or label complexities \cite{nguyen2025applying,he2025automated}. Furthermore, two critical gaps remain: first, a lack of comprehensive comparison evaluating different prompting methods across varied educational contexts and annotation dimensions (i.e., cognitive); and second, a neglect of inherent annotation biases, which need to be addressed to ensure fair and equitable automated analysis.

To address these challenges, this study evaluates the capabilities and biases of LLMs (GPT-5.2 and Gemini-3) in annotating student-AI dialogues. We test their performance using a multi-dimensional scheme across varied contexts, comparing three prompting methods: Few-Shot, Single-agent Self-reflection, and Multi-agent Reflection. We propose two Research Questions (RQs):\textbf{RQ1:} To what extent do prompting methods (few-shot, single-agent, and multi-agent reflection) and contextual factors (educational levels, subjects, and annotation dimensions) influence the accuracy of LLM-based student dialogue annotation? \textbf{RQ2:} What bias patterns emerge in the LLM-based annotation of student dialogues?

\section{Related Work}
\subsection{Educational Dialogue and Student-AI Dialogue Interaction}
Grounded in sociocultural theory, educational dialogue is a critical mechanism for knowledge construction and enhancing academic performance \cite{muhonen2017knowledge,howe2019teacher}. As Large Language Models (LLMs) rapidly popularize student-AI interactions through various chatbots and simulated classrooms \cite{nguyen2025reflective,zhang-etal-2025-simulating}, the lack of human instructor oversight makes the automatic monitoring and decoding of these dialogues essential. Previous research primarily applies framework-based content analysis to evaluate these interactions. Studies have investigated verbal behaviors \cite{zhang-etal-2025-simulating}, cognitive strategies for knowledge construction \cite{dang2025human}, and meta-cognitive behaviors related to AI management \cite{edwards2025human}. Additionally, sentiment analysis is employed to assess the socio-emotional dynamics inherent in student-AI dialogues \cite{dang2025human}.

\subsection{Annotation Based on LLMs and Biases}
LLMs demonstrate significant potential for data annotation across diverse educational levels, subjects, and interactive tasks \cite{hao_mapping_2026,jiang2025uncovering,he2025automated}. However, annotation accuracy varies considerably \cite{jiang2025uncovering}. For example, while GPT-4 achieved high overall agreement with experts in analyzing classroom dialogues, its sub-construct reliability fluctuated widely (Cohen’s Kappa ranging from 0.2 to 0.973) \cite{long2024evaluating}. Such discrepancies stem from differences in models, prompt engineering, and task complexity. The current lack of standardized frameworks necessitates systematic evaluations of LLM performance across various prompting methods and educational contexts to establish a robust understanding of their capabilities. Beyond accuracy, identifying inherent LLM biases is critical for ensuring reliable results. While biases such as human-like cognitive bias (i.e.,  sequential anchoring bias) in decision-making \cite{echterhoff-etal-2024-cognitive}, political bias (i.e., prefer specific political orientation) in detection tasks \cite{lin-etal-2025-investigating}, verbosity bias ( prefer longer text that exceeded word limits) in evaluation \cite{zheng2023judging,qian2025towards} have been documented in other domains, their specific impact on educational dialogue annotation remains underexplored. 

\section{Methodology}
\subsection{Dataset}
Our primary data originates from an online platform where human students interact with AI teachers and peers during slide-based lectures. We collected student utterances from three courses: \textbf{Biology (K-12)} (Fall 2025; 82 students, 304 utterances), \textbf{Introduction to Artificial General Intelligence (AGI, University)} (Spring 2024; 305 students, 4008 utterances), and \textbf{Psychology (University)} (Spring 2025; 288 students, 977 utterances). All data collection procedures were subjected to ethical review by Tsinghua University. Additionally, we incorporated the public \textit{CoMTA} dataset \cite{miller2024llm}, featuring K-12 \textbf{Mathematics} student-AI tutoring conversations. To ensure a balanced evaluation, we randomly sampled 200 student utterances per subject, yielding a final evaluation corpus of 800 utterances (see \href{https://osf.io/yvhar/overview?view_only=eededcccd433490191439605dd9ebd79}{digital appendix}\footnote{\url{https://osf.io/yvhar/overview?view_only=eededcccd433490191439605dd9ebd79}} for samples).

\subsection{Annotation Framework and Human Annotation}
To capture the complexity of student learning, we adopted a four-dimensional annotation framework (detailed rubrics are available in our \href{https://osf.io/yvhar/overview?view_only=eededcccd433490191439605dd9ebd79}{digital appendix}):
\begin{itemize}
    \item \textbf{Behavioral} \cite{han2024recipe4u,hao_mapping_2026}: Seven categories (1=\textit{Request}, 2= \textit{Question}, 3=\textit{Answer}, 4=\textit{Negotiation}, 5=\textit{Acknowledgment}, 6=\textit{Statement}, and 7=\textit{Others/Off-task}).
    \item \textbf{Cognitive} \cite{bloom1956taxonomy}: Four hierarchical levels (0=\textit{None}, 1=\textit{Remembering and Understanding}, 2=\textit{Applying}, and 3=\textit{Analyzing, Evaluating, and Creating}).
    \item \textbf{Meta-cognitive} \cite{zimmerman2002becoming}: Four categories (0=\textit{None}, 1=\textit{Planning and Orientation}, 2=\textit{Monitoring}, and 3=\textit{Reflecting and Evaluating}).
    \item \textbf{Affective} \cite{liu2025engagement}: Six states (1=\textit{Neutral}, 2=\textit{Sense of Accomplishment/Enjoyment}, 3=\textit{Curiosity}, 4=\textit{Frustration/Confusion}, 5=\textit{Anxiety}, and 6=\textit{Boredom}).
\end{itemize}

To establish ground truth, two educational researchers annotated the data. Following two training rounds on a subset ($N=80$), inter-rater reliability achieved a Cohen’s kappa ($\kappa$) $> 0.8$ across all dimensions. The full dataset ($N=800$) was independently double-coded, and any remaining disagreements were resolved via consensus discussions with the first author.

\subsection{Automated Annotations with LLMs}
\subsubsection{Model Selection}
We selected GPT-5.2 (using official default parameters) and Gemini-3 (Flash) (temperature=0.0, max\_output\_tokens=1500) as our experimental models. These represent standard, widely accessible tiers prevalent in educational research, allowing us to assess performance most relevant to cost-efficient, large-scale applications rather than relying on ``Pro'' versions.

\subsubsection{Prompt Engineering Methods}
\begin{figure}
\centering
\includegraphics[width=0.6\textwidth, trim=0cm 0cm 0cm 0cm, clip]{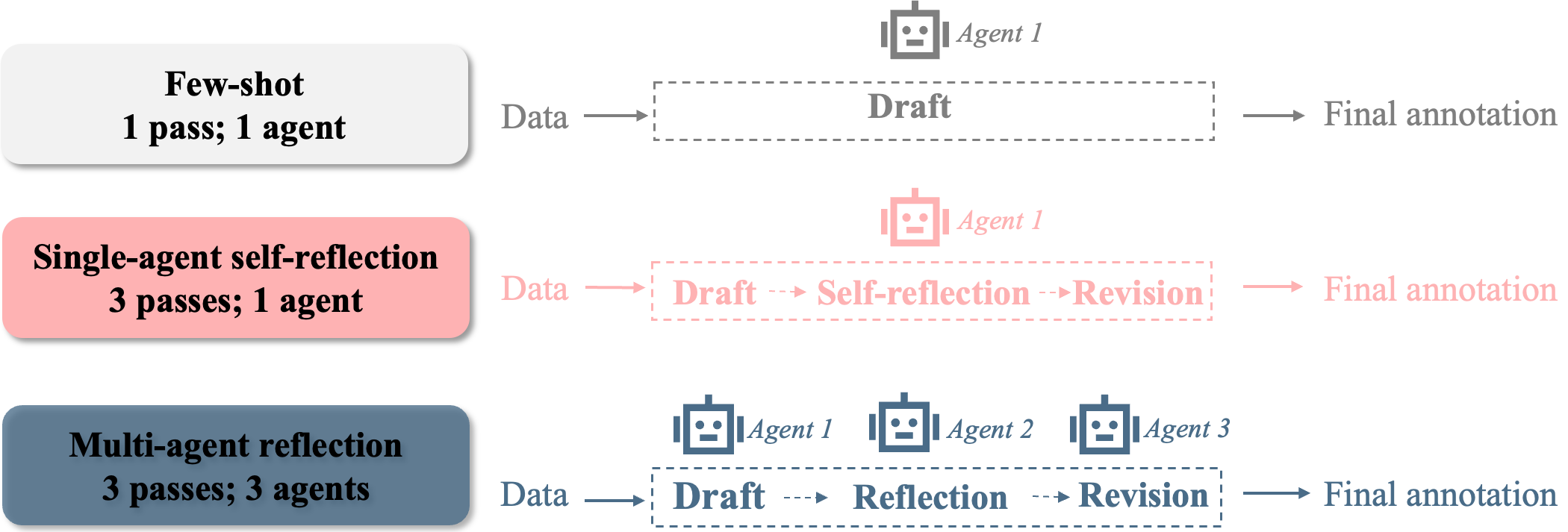}
\caption{The feature of the three adopted prompt engineering methods} \label{fig1}
\end{figure}

To investigate the impact of prompting strategies, we designed three methods (summarized in Figure \ref{fig1}; full prompts available in our \href{https://osf.io/yvhar/overview?view_only=eededcccd433490191439605dd9ebd79}{digital appendix}):
\begin{itemize}
    \item \textbf{Few-shot (Baseline)}: Assigns an educational data expert persona, provides the detailed rubric, and includes three representative examples.
    \item \textbf{Single-agent Self-Reflection} \cite{shinn2023reflexion}: A single agent iteratively generates an initial annotation, self-reflects to identify inconsistencies, and refines its output (Draft $\to$ Self-reflection $\to$ Revision).
    \item \textbf{Multi-agent Reflection} \cite{shinn2023reflexion,cao2025first}: A collaborative system utilizing three specialized agents: Agent 1 drafts the initial annotation, Agent 2 provides a structured critique and reflection, and Agent 3 synthesizes this feedback to produce the final output (Draft $\to$ Reflection $\to$ Revision).
\end{itemize}

\subsection{Data analysis }
To address \textbf{RQ1}, we first calculated the overall annotation Accuracy. Due to perfect collinearity between educational level and subject, we employed a two-step Generalized Linear Mixed Model (GLMM) strategy to analyze utterance-level annotation correctness (binary: correct/incorrect). All models included random intercepts for RowID to account for item-level variability. In the first step, we constructed a global GLMM including prompt method, educational level, and annotation dimension as fixed effects (excluding subject), followed by post-hoc pairwise comparisons for these variables. In the second step, we conducted separate subgroup GLMM analyses within each educational level to evaluate subject-specific effects.

To address \textbf{RQ2}, we defined several indicators to quantify biases between LLM predictions and human annotations and report the percentage:
\begin{itemize}
    \item \textbf{Affective Bias}: Affective Optimistic Bias (AOB): The proportion of instances where the predicted affective value was more positive than the human-annotated results (e.g., predicting ``Neutral'' as ``Curiosity''). Affective Pessimistic Bias (APB):  The proportion of instances where the predicted sentiment polarity was more negative than the human-annotated results.
    \item \textbf{Cognitive Bias}: Cognitive Overestimation Bias (COB): The proportion of instances where the predicted cognitive level was at least one unit higher than the human-annotated results; Cognitive Underestimation Bias (CUB): The proportion of instances where the predicted cognitive level was at least one unit lower than the human-annotated results.
    \item \textbf{Meta-cognitive Bias}: Meta-cognitive Overestimation Bias (MOB): The proportion of instances where non-meta-cognitive behavior was misclassified as a meta-cognitive state. Meta-cognitive Underestimation Bias (MUB):  The proportion of instances where meta-cognitive behavior was misclassified as a non-meta-cognitive state. 
    \item \textbf{Behavioral Bias}: We utilized Sankey diagrams to conduct an exploratory analysis of bias patterns across behavioral categories.
\end{itemize}

\begin{figure}
\centering
\includegraphics[width=0.5\textwidth, trim=0cm 0cm 0cm 0cm, clip]{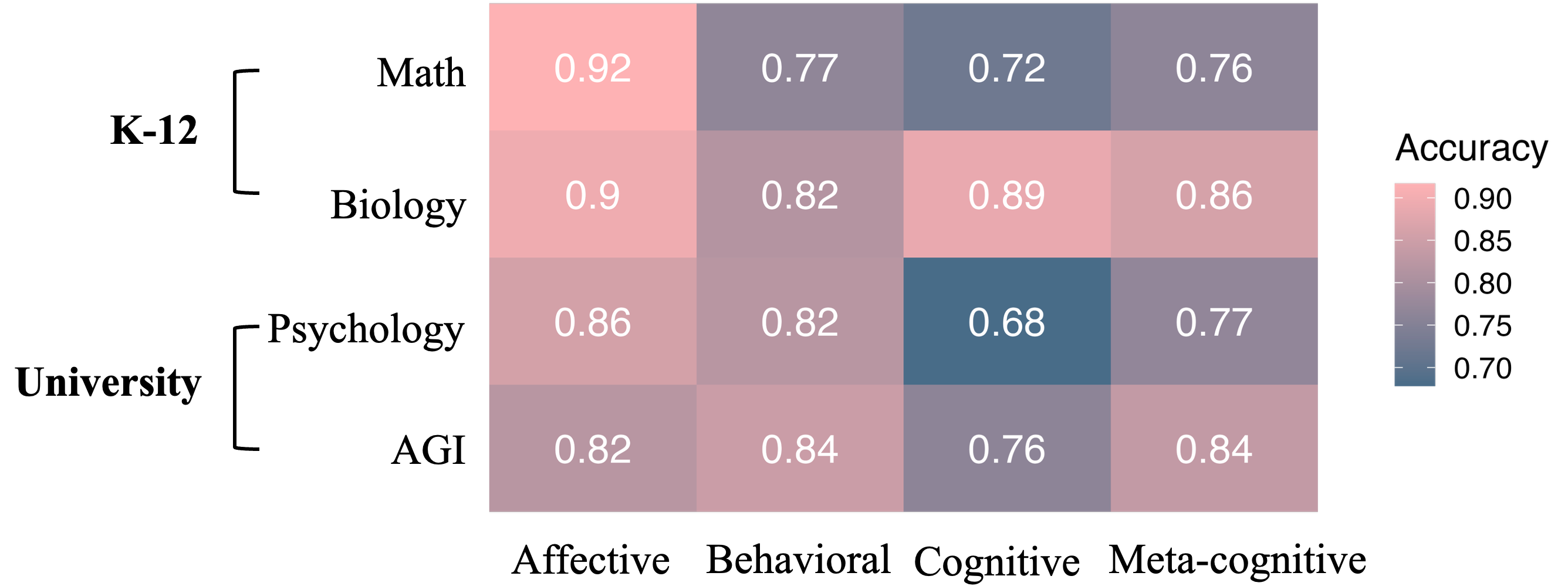}
\caption{Annotation accuracy by subjects} \label{fig3}
\end{figure}

\section{Results}
\subsection{Contextual and Dimensional Variations in Accuracy, but Not Prompting Methods}
\begin{table}[htbp]
\small
\centering
\caption{Fixed effects of educational level, dimension on annotation accuracy}\label{tab:model_results}
\footnotesize %
\renewcommand{\arraystretch}{1.0} %
\setlength{\tabcolsep}{5pt} %
\begin{tabular}{lccc}
\toprule
\textbf{Variable} & \textbf{Estimate} & \textbf{Std. Error} & \textbf{$p$-value} \\
\midrule
(Intercept) & 1.993 & 0.072 & < 0.001 *** \\
\addlinespace[0.2em] %

\textbf{Model (Ref: Gemini-3)} & & & \\
GPT-5.2 & 0.265 & 0.038 & < 0.001 *** \\
\addlinespace[0.2em]

\textbf{Method (Ref: Few-shot)} & & & \\
Single-agent Self-reflection& 0.047 & 0.046 & 0.314 \\
Multi-agent Reflection& 0.071 & 0.047 & 0.127 \\
\addlinespace[0.2em]

\textbf{Educational Level (Ref: K-12)} & & & \\
University & $-$0.212 & 0.038 & < 0.001 *** \\
\addlinespace[0.2em]

\textbf{Dimension (Ref: Affective)} & & & \\
Behavior & $-$0.487 & 0.058 & < 0.001 *** \\
Cognitive & $-$0.796 & 0.056 & < 0.001 *** \\
Meta-cognitive & $-$0.518 & 0.058 & < 0.001 *** \\

\bottomrule
\end{tabular}
\end{table}

Descriptively, annotation accuracy slightly improved as prompt complexity increased: from the Few-Shot (Gemini-3: 79.2\%; GPT-5.2: 82.4\%) to Single-Agent (79.5\% and 83.4\%) and Multi-Agent (79.7\% and 83.9\%). However, GLMM results (Table \ref{tab:model_results}) revealed these marginal enhancements were not statistically significant; neither Single-agent Self-reflection ($p = .314$) nor Multi-agent Reflection ($p = .127$) significantly outperformed the Few-shot. In contrast, contextual factors significantly impacted the likelihood of correct annotation. University-level dialogues proved significantly more difficult to annotate than K-12 dialogues ($b = -0.212, p < .001$). Subgroup analyses within educational levels showed that correctness in Mathematics was significantly lower than in Biology (K-12, $p < .001$), whereas AGI outperformed Psychology (University, $p < .001$). Finally, significant variations emerged across annotation dimensions. Compared to the Affective, models demonstrated significantly lower correctness on Behavioral, Meta-cognitive, and Cognitive dimensions (all $p < .001$), with the Cognitive dimension proving the most challenging ($b = -0.796$). Detailed accuracies across dimensions and subjects are illustrated in Figure \ref{fig3}.

\subsection{Annotation Bias}
\begin{figure}
\centering
\includegraphics[width=\textwidth, trim=0cm 0cm 0cm 0cm, clip]{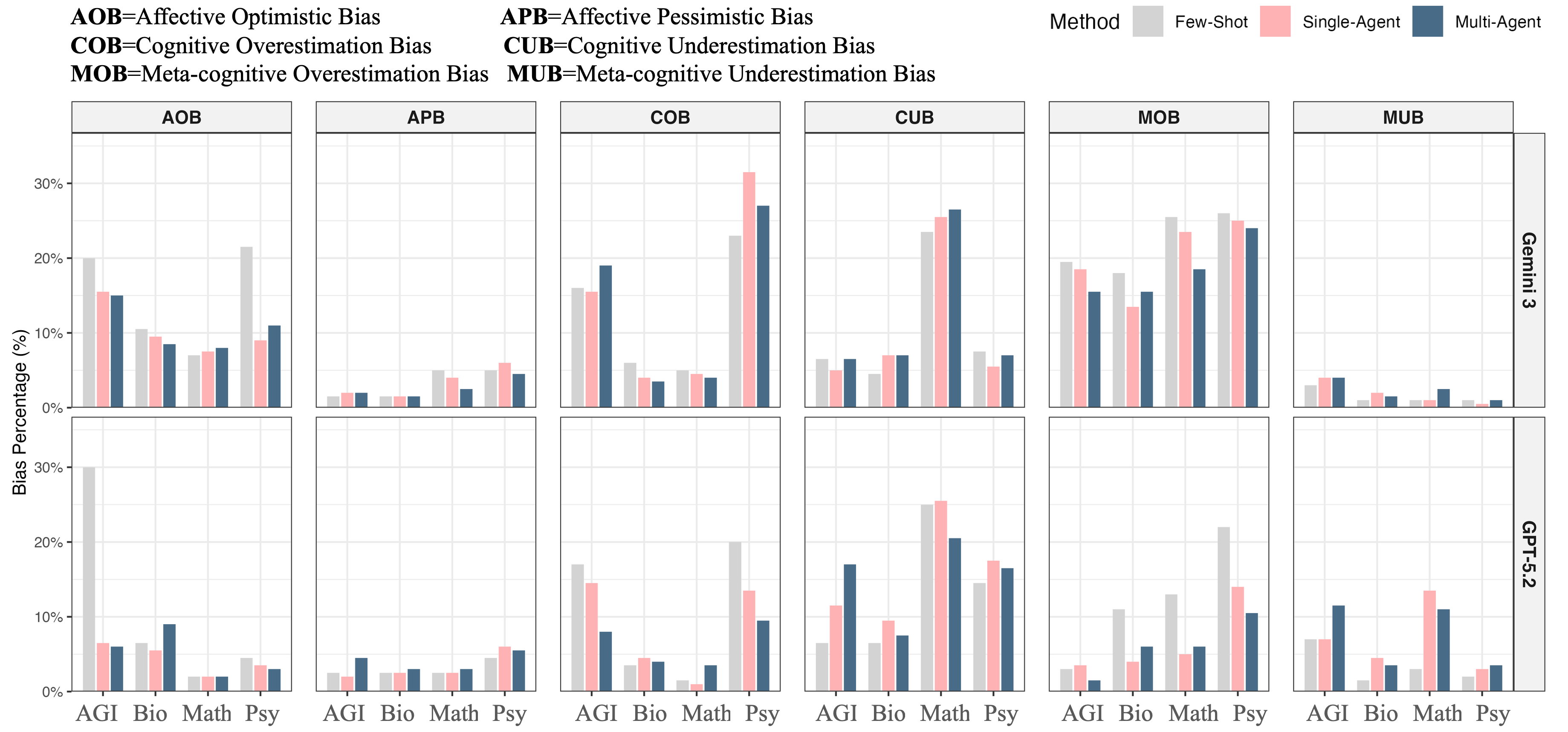}
\caption{The annotation bias (affective, cognitive, meta-cognitive) of LLMs} \label{fig4}
\end{figure}

As illustrated in Figure \ref{fig4}, we observed distinct bias patterns across the four dimensions:(1)\textbf{Affective Bias}: Gemini-3 exhibited a substantially higher Affective Optimistic Bias (AOB) than GPT-5.2 across most subjects, frequently over-attributing positive emotions, while their Pessimistic Bias (APB) remained comparable. For example, in Psychology, Gemini-3's AOB ranged from 9\% to 21.5\% across prompt methods, compared to GPT-5.2's 3\% to 4.5\%. A notable exception occurred in AGI, where GPT-5.2 (Few-shot) showed a higher AOB (30\%) than Gemini-3 (20\%).
(2)\textbf{Cognitive Bias}: Mathematics uniquely suffered from Cognitive Underestimation Bias (CUB) across both models and all prompts, peaking at 26.5\% for Gemini-3 (Multi-agent). Conversely, Cognitive Overestimation Bias (COB) was prominent in AGI and Psychology, but remained minimal in the STEM disciplines (Biology and Mathematics).
(3)\textbf{Meta-cognitive Bias}: Gemini-3 consistently displayed higher Meta-cognitive Overestimation Bias (MOB), frequently misclassifying non-meta-cognitive utterances. In Psychology, Gemini-3's MOB ranged from 24\% to 26\%, vastly exceeding GPT-5.2's 0.5\% to 1\%. While GPT-5.2 showed a slightly higher Underestimation Bias (MUB) than Gemini-3, overall MOB rates consistently exceeded MUB rates for both models.
(4)\textbf{Behavioral Bias}: LLMs frequently misclassified human-labeled Questions, Negotiations, and Statements. In Mathematics specifically, nuanced interactions caused confusion: Questions were often mislabeled as Negotiations or Acknowledgments, and Negotiations as Requests or Questions (See Sankey diagram in the \href{https://osf.io/yvhar/overview?view_only=eededcccd433490191439605dd9ebd79}{Digital Appendix}). However, the models aligned well with human annotators on explicit categories like Requests and Off-task behaviors.

\section{Discussion and Conclusion}
This study evaluates LLMs (Gemini-3 and GPT-5.2) as automated annotators for student-AI educational dialogues. Our results indicate that increasing prompt complexity yielded no statistically significant improvements over the baseline. Consequently, the few-shot approach remains a cost-effective alternative in resource-constrained environments. Crucially, our findings indicate that annotation accuracy is inherently context-dependent and dimension-sensitive, resonating with other studies \cite{jiang2025uncovering}. LLMs performed better on K-12 dialogues than University-level interactions. Performance also varied by subject (e.g., AGI outperformed Psychology), likely due to the epistemological nature of domains like Psychology, where subjective and open-ended knowledge complicates the assessment \cite{zheng2023judging}. We also identified a performance hierarchy across dimensions: LLMs excelled in the affective dimension—consistent with established strengths in sentiment analysis \cite{annan2025evaluating}—but struggled with the cognitive dimension \cite{long2024evaluating,he2025automated}. Moving beyond standard accuracy metrics, this research characterizes systematic biases inherent in LLM annotations, echoing observations in other domains \cite{lin-etal-2025-investigating,echterhoff-etal-2024-cognitive}. Notable patterns included a pronounced AOB in Gemini-3, domain-specific cognitive biases (CUB in Mathematics; COB in Psychology), and a systemic tendency toward MOB in the meta-cognitive dimension. These findings suggest that future prompts need to be tailored to specific models, disciplinary characteristics, and annotation dimensions to effectively mitigate bias. Furthermore, frequent conflation of behavioral categories suggests that single utterances often serve multiple pragmatic functions. Researchers should thus ensure strict mutual exclusivity in taxonomies or adopt multi-label classification frameworks.

In conclusion, while LLMs demonstrate considerable potential for scaling educational dialogue analysis, their reliability is bounded by contextual nuances and systemic biases. Given the limitations in current study, future studies should incorporate more advanced LLMs and evaluate more subjects strictly controlled across educational levels to facilitate a holistic assessment of systematic biases. Furthermore, to effectively mitigate these biases to build a more fair LLM-based annotation, beyond  prompt adjustments, further study should pursue targeted model fine-tuning and establish robust human-AI collaborative workflows.

\begin{credits}
\subsubsection{\ackname}  This work was supported by the National Natural Science Foundation of China (No.62407027) and the Beijing Educational Science Foundation of the Fourteenth 5-year Planning (BAEA24024).
\end{credits}

%
%
%
\bibliographystyle{splncs04}
\bibliography{Ref}
%





\end{document}